\documentclass[pra,twocolumn,floatfix,superscriptaddress]{revtex4-1}
\usepackage{dcolumn,amsmath}
\usepackage{graphicx}
\usepackage{bm}
\usepackage{amssymb}
\usepackage{multirow}

\usepackage{amsfonts}
\usepackage{amsmath}
\usepackage{cancel}
\usepackage{xcolor}
\newcommand{\ecm}{\ensuremath{e {\cdotp} {\rm cm}}}

\newcommand{\de}{d_\mathrm{e}}

\begin{document}
\title{Electronic matrix elements for parity doubling in YbOH molecule}
\author {Alexander Petrov}\email{petrov\_an@pnpi.nrcki.ru}
\affiliation{Petersburg Nuclear Physics Institute named by B.P. Konstantinov of National Research Centre
"Kurchatov Institute", Gatchina, 1, mkr. Orlova roshcha, 188300, Russia}
\affiliation{St. Petersburg State University, St. Petersburg, 7/9 Universitetskaya nab., 199034, Russia} 

\date{Received: date / Revised version: date}
%
\begin{abstract}{
 YbOH molecule is one of the most sensitive systems for the electron electric dipole moment ($e$EDM) searches. The $e$EDM-induced energy shift is proportional to polarization ($P$) of the molecule.
 In Ref. [A. Petrov and A. Zakharova, Phys. Rev. A 105,
L050801 (2022)]  was shown that the
value of l-doubling and spin-rotation splitting directly influences the maximum value of $P$.
Recently in Ref. [Jadbabaie, Y. Takahashi, N. H. Pilgram, C. J. Conn, Y. Zeng, C. Zhang, and N. R. Hutzler, New Journal of
Physics 25, 073014 (2023)] the corresponding energy levels was determined experimentally. We introduced electronic matrix elements in Hund's case $c$ coupling scheme to reproduce experimental energy levels and calculated $P$ as function of external electric field.
} 
\end{abstract}
\maketitle
\section{Introduction}
Measuring the electron electric dipole moment ($e$EDM) is now considered as a most promising test for existence of physics beyond the Standard model \cite{Fukuyama2012,PospelovRitz2014,YamaguchiYamanaka2020,YamaguchiYamanaka2021}. 
The current constrain for $e$EDM $|\de|<4.1\times 10^{-30}$ \ecm\ (90\% confidence) was obtained using trapped $^{180}$Hf$^{19}$F$^+$ ions \cite{newlimit1} with spinless $^{180}$Hf isotope. Cold polar triatomic molecules provide opportunities for further progress in search for effects of symmetry violation \cite{Isaev:16}.
In such molecules the sensitivity of the experiments can be strongly enhanced due to laser cooling \cite{Isaev_2017} and the existence of  
$l$-doublets of the excited $v=1$ bending vibrational mode helps to suppress many sistematics \cite{Kozyryev:17,Pilgram:21}. 

A great progress is achieved recently in both theoretical and experimental study of triatomics. In Ref. \cite{anderegg2023quantum} a quantum control of trapped triatomic molecules  for $e$EDM searches is demonstrated. In Ref. \cite{Jadbabaie_2023} a detailed spectroscopy of the $e$EDM sensitive $l$-doublets of the ground rotational $N=1$ level of the excited $v=1$ bending vibrational mode of $^{174}$YbOH was performed. Unusually large, compared to other metal
hydroxides, asymmetric parity-doubling of the $J = 1/2$ and $J = 3/2$ manifolds was revealed. As is noted in Ref. \cite{Jadbabaie_2023} a further study is required to determine in details the nature of this asymmetry. 
In Ref. \cite{Petrov:2022} the method for computation of energy levels and different properties of triatomic molecules was developed. The method was applied for calculation of sensitivity of the $^{174}$YbOH molecule to $e$EDM in the ground rotational $N=1$ level of first excited $v=1$ bending mode in the external electric field. In calculations (see below for details) matrix elements for $v=1$ 
were assumed to be equal to ones for $v=0$ and taken from Ref. \cite{Pilgram:21}. In this approximation there were no asymmetry in parity-doubling of the $J = 1/2$ and $J = 3/2$. 
In Ref.  \cite{Petrov:2022}  we have shown that the
value of l-doubling and spin-rotation splitting directly influences the maximum degree of $T,P$-odd polarization and thus influences the sensitivity of linear triatomic molecules to the $T,P$-odd effects.
Therefore,
in the current work we modiied and introduced new {\it electronic} matrix elements in Hund's case $c$ coupling scheme, which allows us to reproduce experimental energy levels and in particular asymmetry of the $J = 1/2$ and $J = 3/2$ manifolds. As opposite to Hund's case $b$ using Hund's case $c$ coupling scheme helps in future to calculate this effect {\it ab initio} as many modern quantum chemical packages allow to include the spin-orbit interaction in all orders. Using newly obtained electronic matrix elements we recalculated the sensitivity of the $^{174}$YbOH molecule to $e$EDM in the external electric field.

\section{Method} 
Following Ref. \cite{Petrov:2022}  we present our Hamiltonian in molecular reference frame as
\begin{equation}
{\rm \bf\hat{H}} = {\rm \bf\hat{H}}_{\rm mol} + {\rm \bf\hat{H}}_{\rm hfs} + {\rm \bf\hat{H}}_{\rm ext},
\label{Hamtot}
\end{equation} 
where
\begin{equation}
{\rm \bf\hat{H}}_{\rm mol}=\frac{(\hat{\bf J} -\hat{\bf J}^{e-v} )^2}{2\mu R^2}+\frac{(\hat{\bf J}^{v})^2}{2\mu_{\rm OH}r^2}+ V(\theta)
\label{Hmolf}
\end{equation}
is the molecular Hamiltonian,
$\mu$ is the reduced mass of the Yb-OH system, $\mu_{\rm OH}$ is the reduced mass of the OH, $\hat{\bf J}$ is the total electronic, vibrational, and rotational
angular momentum, $\hat{\bf J}^{e-v} = \hat{\bf J}^{e} + \hat{\bf J}^{v}$ is the electronic-vibrational momentum, $\hat{\bf J}^{e}$ is the electronic momentum, $\hat{\bf J}^{v}$ is the vibrational momentum,
$R$ is the distance between Yb and the center mass of OH, $r$ is OH bond length
and $\theta$ is the angle between OH  and the axis ($z$ axis of the molecular frame) directed from Yb to the OH center of mass. The condition $\theta=0$ corresponds to the linear configuration where the O atom is between Yb and H ones. $R$, $r$ and $\theta$ are the so called Jacobi coordinates.

 In the current work we have fixed $R$ and $r$ in such a way that
 $\frac{\hbar^2}{2\mu R^2} = 7329$ MHz to reproduce the experimental value for rotation constant \cite{Jadbabaie_2023} and $\frac{\hbar^2}{2\mu_{\rm OH} r^2} = 19.6$ cm$^{-1}$ to fit experimental value of 24 MHz for $l-$doubling \cite{Jadbabaie_2023}. Recently the same value was also obtained in {\it ab initio} calculations \cite{Zakharova:22}.
In this approximation we neglect the influence of the stretching (associated with R) and OH ligand (associated with r)  modes but nevertheless
take into account the bending ones (associated with $\theta$) with fixed $R,r$.
$V(\theta)$ is the potential energy curve obtained in the electronic structure calculations \cite{Zakharova:21b}.

$ {\rm \bf\hat{H}}_{\rm hfs} $  and  ${\rm \bf\hat{H}}_{\rm ext}$ are the hyperfine interaction with H nucleus and Stark interaction with the external electric field respectively, as they are described in Ref.~\cite{Petrov:2022}

Wavefunctions, rovibrational energies and hyperfine structure were obtained by numerical diagonalization of the Hamiltonian (\ref{Hamtot})
over the basis set of the electronic-rotational-vibrational-nuclear spins wavefunctions
\begin{equation}
 \Psi_{\Omega }P_{lm}(\theta)\Theta^{J}_{M_J,\omega}(\alpha,\beta)U^{\rm H}_{M^{\rm H}_I}.
\label{basis}
\end{equation}
Here 
 $\Theta^{J}_{M_J,\omega}(\alpha,\beta)=\sqrt{(2J+1)/{4\pi}}D^{J}_{M_J,\omega}(\alpha,\beta,\gamma=0)$ is the rotational wavefunction, $\alpha,\beta$ correspond to azimuthal and polar angles of the $z$ axis,
 $U^{\rm H}_{M^{\rm H}_I}$ is the hydrogen  nuclear spin wavefunction, $M_J$ is the projection of the molecular (electronic-rotational-vibrational) angular momentum $\hat{\bf J}$ on the lab axis, 
 $\omega$ is the projection of the same momentum on $z$ axis of the molecular frame,
 $M^{\rm H}_I$  is the projections of the nuclear angular 
momenta of hydrogen  on the lab axis, $P_{lm}(\theta)$ is the associated Legendre polynomial,
$l$ is the vibration angular momentum and $m$ is its projection on the molecular axis, 
$\Psi_{\Omega}$ is the electronic wavefunction (see Ref. \cite{Petrov:2022} for details).

In this  calculation functions with $\omega - m = \Omega = \pm 1/2$, $l=0-30$  and $m=0,\pm 1, \pm 2$, $J=1/2,3/2,5/2$  were included to the basis set (\ref{basis}).
The ground vibrational state $v=0$ corresponds to $m=0$,
the first excited bending mode $v=1$ (focus of this paper) to $m=\pm 1$, the second excited bending mode $v=2$ has states with $m=0, \pm2$ etc. 

Provided that the {\it electronic-vibrational} matrix elements are known, the matrix elements of ${\rm \bf\hat{H}}$ between states in the basis set (\ref{basis}) can be calculated with help of the angular momentum algebra \cite{LL77, Petrov:2022} mostly in the same way as for the diatomic molecules \cite{Petrov:11}.

The required matrix elements associated with H nucleus magnetic hyperfine interaction were taken from Ref. \cite{Petrov:2022}. Dipole moment operator 
\begin{equation}
   \langle
   \Psi_{\Omega}  | 
   D_z
|\Psi_{\Omega}\rangle 
= -0.850
{~ \rm a.u.}
\label{dipvalue}
\end{equation}
 determining interaction with the external electric field was taken from Ref. \cite{Jadbabaie_2023}.
 
A special attention is given for the matrix element of $J_+^e = J_x^e + iJ_y^e$ operator which, in particular, ensure asymmetry of the $l-$doubling of $J = 1/2$ and $J = 3/2$ manifolds. We put
\begin{multline}
\frac{1}{\mu R^2}
   \langle\Psi_{\Omega=1/2} |J^e_+|\Psi_{\Omega=-1/2} \rangle 
= p_0 + p_1P_{l=1m=0}(\theta),
\label{pme}
\end{multline}
\begin{multline}
\frac{1}{\mu R^2}
   \langle
   \Psi_{\Omega=-1/2} |J^e_+
|\Psi_{\Omega=+1/2} \rangle 
= p_2P_{l=2m=2}(\theta).
\label{pme2}
\end{multline}
Here we take into account that pure electronic matrix elements, in general, depend on $\theta$,
and the selection rules for $\Omega$ quantum number can be violated \cite{Petrov:2022}.
It is assumed that $\Psi_{\Omega}$ are chosen in such a way that  $\langle\Psi_{\Omega=1/2} |\partial / \partial \theta|\Psi_{\Omega=-1/2} \rangle = 0$.
Note that here $\Omega$ is the projection of the total electronic angular momentum on the molecular axis $z$ for {\it linear} configuration.
Equation $\hat{ J}^{e}_z \Psi_{\Omega }(\theta) = \hbar \Omega \Psi_{\Omega }(\theta)$ is not satisfied
for the bending configuration $\theta \ne 0$.  
The selection rules for $\omega$ quantum number are rigorous and are the same as for $\Omega$ quantum number for the linear configuration.

\section{Results}
\subsection{Energy levels for field-free case}
\begin{figure}
\includegraphics[width=1.0\linewidth]{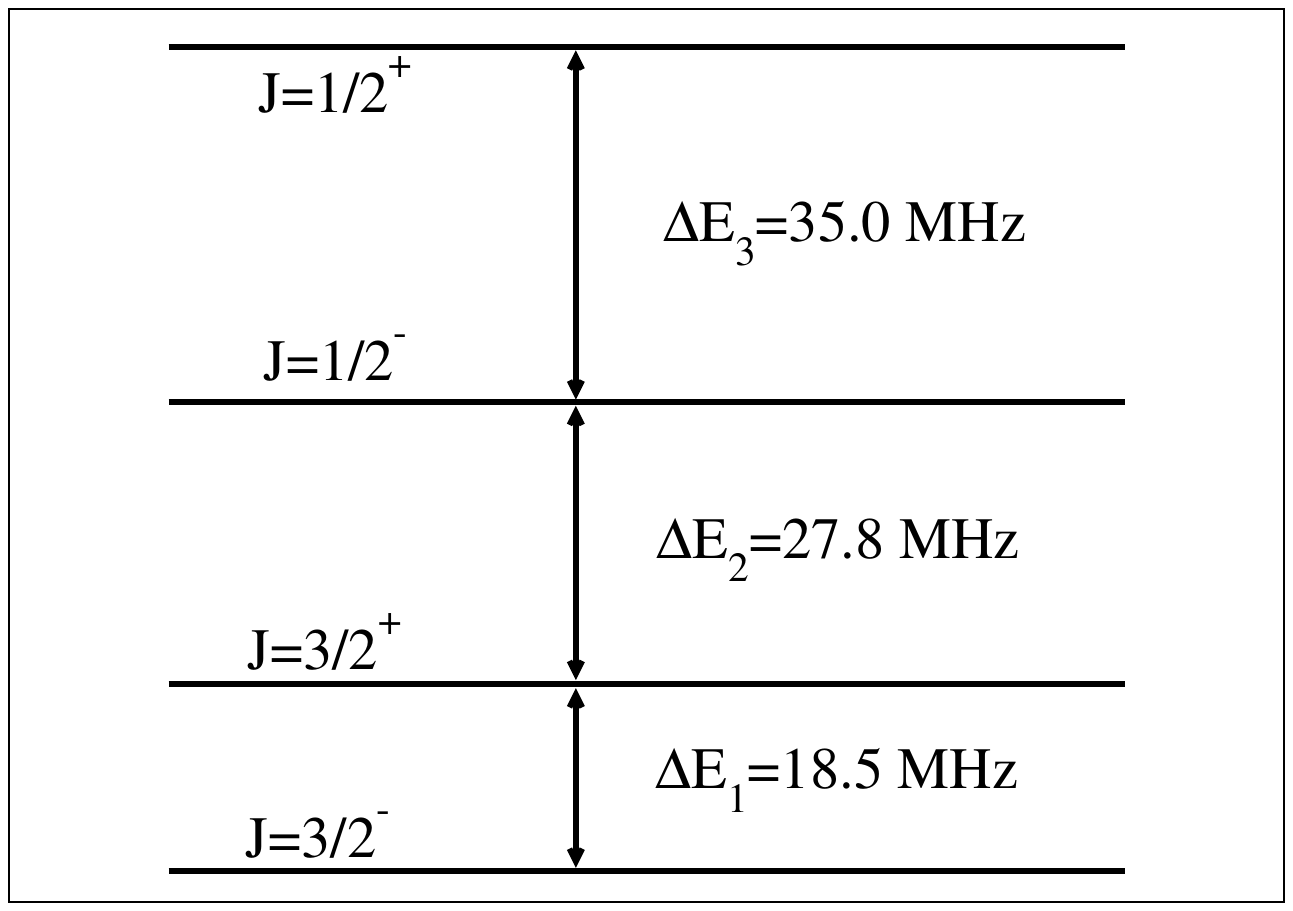}
\caption{\label{elevels} 
Experimental energies of the ground rotational $N=1$ level of the excited $v=1$ bending vibrational modes of $^{174}$YbOH \cite{Jadbabaie_2023}. Parity of the states is shown as a superscript. Unusually large asymmetry of parity-doubling of the $J = 1/2$ ($\Delta E_3$) and $J = 3/2$ ($\Delta E_1$) manifolds is observed. Hyperfine structure is not shown.}
\end{figure}

The rotational levels of the excited $v=1$ bending vibrational mode of YbOH are well described by the Hund's case $b$ coupling scheme \cite{Jadbabaie_2023}. Electron spin ${ S=1/2}$ for a good approximation is an integral of motion. Its interaction (spin-rotation) with the rovibrational momentum ${\bf N} = {\bf J} - {\bf S}$ gives rise to the splitting between the energy levels with total $J = N \pm 1/2$ momenta. Each level has two parity eigenstates -- l-doublets. $l-$doubling is, in general, different for the $J = N + 1/2$ and $J = N - 1/2$ levels. The experimental energy levels obtained in Ref. \cite{Jadbabaie_2023} for the ground rotational level $N=1$ are depicted in Fig. \ref{elevels}.

 The $p_0$ value used in the Hund's case $c$ coupling scheme can be obtained by equation \cite{Kozlov:95} 
 \begin{equation}
 p_0 = \frac{\hbar^2}{\mu R^2} -\gamma = 0.492  {\rm ~cm}^{-1},
 \label{p0vsg}
 \end{equation}
 where $\gamma = -88.7$ 
 MHz determines spin-rotation interaction $\gamma (\hat{\bf S} \cdot \hat{\bf N} -S_zN_z)$ in the Hund's case $b$ coupling scheme \cite{Jadbabaie_2023}. Use the correlated by Eq. (\ref{p0vsg}) $p_0$ and $\gamma$ gives the same energy levels in both Hund's case coupling schemes. In this approximation (corresponds to the calculations in Ref. \cite{Petrov:2022}) $l-$doubling is the same for  $J = N \pm 1/2$ levels ($\Delta E_1 \approx \Delta E_3$, see Table \ref{deres} and Fig. \ref{elevels} for definition of $\Delta E_i$). To reproduce the experimental energy levels we also put 
 \begin{equation}
 p_1 =  -37.0  {\rm ~MHz},
 \label{p1}
 \end{equation}
 \begin{equation}
 p_2 =  125.9   {\rm ~MHz}.
 \label{p2}
 \end{equation}
 In Table \ref{deres} the calculated values $\Delta E_1$, $\Delta E_2$ and $\Delta E_3$.
 for the cases (A) $p_0 = 0.492  {\rm ~cm}^{-1}$,  $p_1 =  0.0  {\rm ~MHz}$, $p_2 =  0.0   {\rm ~MHz}$; (B) $p_0 = 0.492  {\rm ~cm}^{-1}$,  $p_1 =  0.0  {\rm ~MHz}$, $p_2 =  125.9   {\rm ~MHz}$; (C) $p_0 = 0.492  {\rm ~cm}^{-1}$,  $p_1 =  -37.0  {\rm ~MHz}$, $p_2 =  125.9   {\rm ~MHz}$ (corresponds to optimal matrix elements (\ref{p0vsg} -\ref{p2})) are given.

 \begin{table}
\caption{ The calculated $\Delta E_1$, $\Delta E_2$ and $\Delta E_3$ energy splittings (MHz) for the $p_0, p_1, p_2$ parameters corresponding to the (A), (B) and (C) cases. In (A) case only $p_0$ constant is taken into account. In (B) case $p_1$ constant is added. In (C) case all three constants are taken into account.
See text for details. Case (C) reproduces experimental values for
 $\Delta E_1$, $\Delta E_2$ and $\Delta E_3$. In brackets for the (B) case the increment from the (A) case, for the (C) case the increment from the (B) case are given.} 
 
\label{deres}
\begin{tabular}{cccc}
                     & (A) & (B) & (C) \\
\hline
$\Delta E_1$ & 23.9 &  18.5 (-5.4) &  18.5 (0.0)\\
$\Delta E_2$ & 42.4 &  39.7 (-2.7) &  ~~27.8 (-11.9) \\
$\Delta E_3$ & 24.1 &  35.0 (10.9) &  35.0 (0.0)\\
\end{tabular}
\end{table}

Accounting for the $p_2$ constant leads to the asymmetry of the $l-$doubling of $J = 1/2$ ($\Delta E_3$) and $J = 3/2$ ($\Delta E_1$) manifolds.
One can see that increments ($\delta \Delta E_i$) for $\Delta E_i$ energy splittings when $p_2$ constant is taken into account have ratios $\delta \Delta E_1 / \delta \Delta E_2 = 2$ and $\delta \Delta E_3 / \delta \Delta E_2 = -4$. Exactly the same ratios are for the Hamiltonian $p_G/2(N_+S_+e^{-i2\phi} + N_-S_-e^{+i2\phi})$ used in Ref. \cite{Jadbabaie_2023} for the Hund's case $b$ coupling scheme. Therefore one should associate $p_G$ and $p_2$ constants. From Eq. (\ref{pme2}) one can see that $p_2$ constant is nonzero when $\Omega$ quantum number is violated for the bending configuration.

Accounting for the $p_1$ constant leads only to the increment of $\Delta E_2$. The same effect is for the Hamiltonian $\gamma_GN_zS_z$ used in Ref. \cite{Jadbabaie_2023}. Therefore one should associate $\gamma_G$ and $p_1$ constants. From Eq. (\ref{pme}) one can see that $p_1$ constant can be nonzero without violation of the $\Omega$ quantum number.

To the best of our knowledge, calculation of constants $p_1$ and $p_2$ is not currently available in public quantum-chemical codes and should be a goal for the further development.

 \subsection{Sensitivity to the $e$EDM}

Any $e$EDM experiment searches for an $e$EDM induced Stark shift
\begin{equation}
\delta E = P E_{\rm eff} d_e,
\label{split}
\end{equation}
where $d_e$ is the value of electron electric dipole moment, $E_{\rm eff}$ is {\it effective electric field} acting on electron in the molecule,
$P$ is the polarization of the molecule by the external electric field.
 (We note, that $P$ is not equal to the mean value of the projection of unit vector $\hat{z}$ along molecular axis on direction of the external electric field.)
To extract $d_e = \delta E / (E_{\rm eff} P) $ from the measured
shift $\delta E $, one needs to know $PE_{\rm eff}$.
$E_{\rm eff}$ was a subject of molecular calculations
\cite{denis2019enhancement,prasannaa2019enhanced,gaul2020ab, Zakharova:21b}.

\begin{figure}
\includegraphics[width=0.95\linewidth]{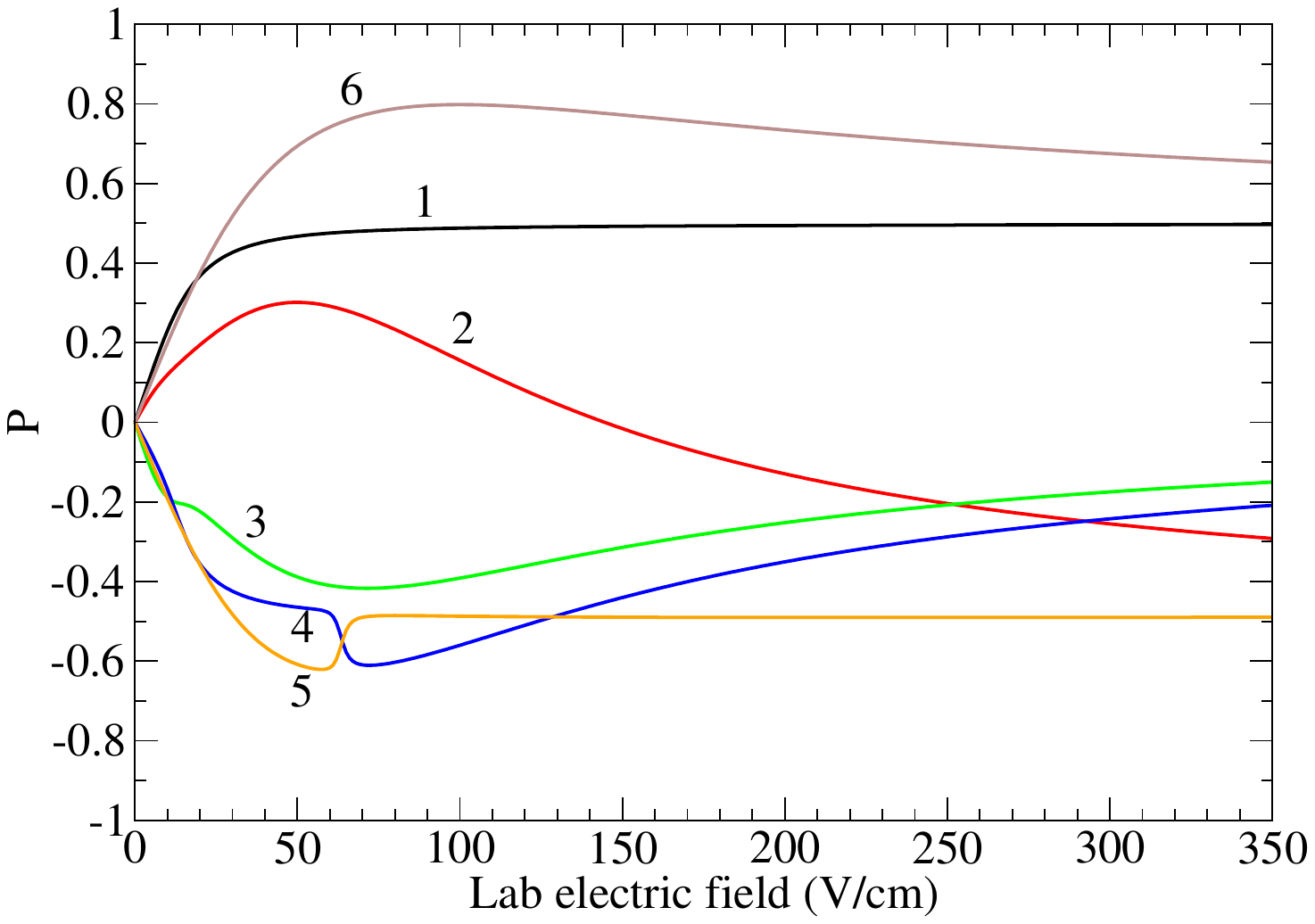}
\caption{\label{EDMshift} 
(Color online) Calculated polarization $P$ (see eq. (\ref{split})) for the  $M_F=M_J+M_I=1$ hyperfine sublevels of the lowest $N=1$ rotational level
of the first excited the $v=1$ bending vibrational mode of $^{174}$YbOH as function of the external electric field. Colors (numbering) of lines correspond to colors (numbering) of lines in Fig. \ref{Ener}.}

\end{figure}

\begin{figure}
\includegraphics[width=0.95\linewidth]{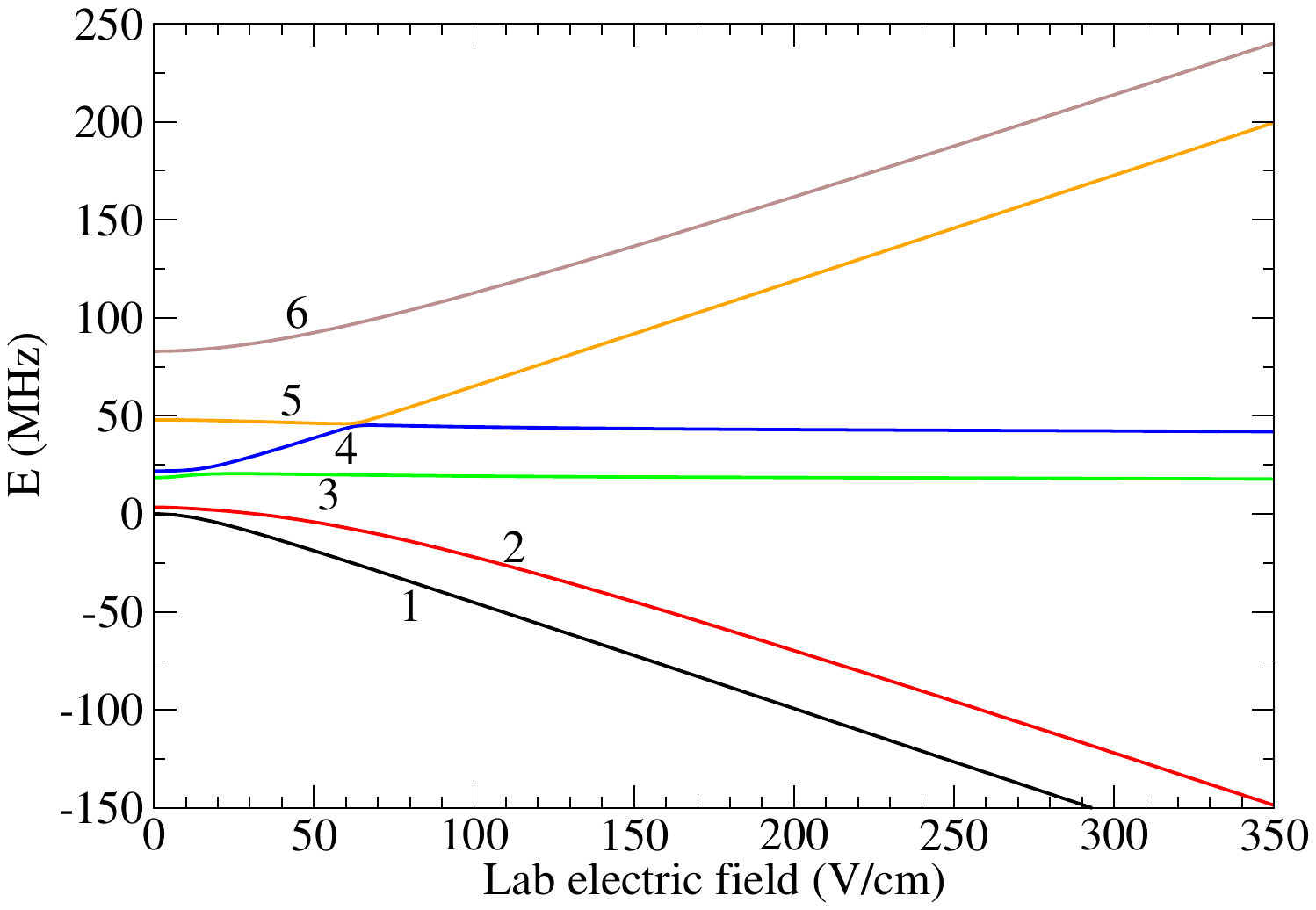}
\caption{\label{Ener} 
(Color online) Calculated energies for the  $M_F=M_J+M_I=1$ hyperfine sublevels of the lowest $N=1$ rotational level
of the first excited $v=1$ bending vibrational mode of $^{174}$YbOH as functions of the external electric field.
Colors (numbering) of lines correspond to colors (numbering) of lines in Fig. \ref{EDMshift}.}
\end{figure}

In this work to calculate $P$ we also include hyperfine interaction with the hydrogen nucleus.
Hydrogen nucleus has a nonzero nuclear spin I = 1/2, which gives rise to
the hyperfine energy splitting between the levels with total
(electronic-vibrational-rotational-nuclear spin) angular momentum $F = J \pm 1/2$
(not showed on Fig. \ref{elevels}).  In Fig. \ref{EDMshift} the calculated
polarizations $P$ for six $M_F = M_J + M_I = 1$ hyperfine sublevels of the lowest N = 1 rotational level of
the first excited $v = 1$ bending vibrational mode of the
$^{174}$YbOH as function of the external electric field are given.
The maximum of polarization $P=0.80$ is reached for the sixth level at electric field 100 V/cm in agreement with data of Ref. \cite{Jadbabaie_2023}. In Fig. \ref{Ener} the corresponding calculated energy levels are presented.

\section{Conclusion}
We determined electronic matrix element 
in Hund's case $c$ coupling scheme to reproduce experimental energy levels and in particular asymmetry in $l-$doubling structure of the ground rotational level of the first excited bending vibrational mode of $^{174}$YbOH. Matrix elements can be associated with the parameters of the effective Hamiltonian in Hund's case $b$ coupling scheme. $T,P-$odd polarization determining sensitivity to $e$EDM is calculated as function of the external electric field. The maximum value $P=$0.8 is found for electric field $E=100$ V/cm.
%
%

\end{document}